\documentclass[10pt,showpacs,preprintnumbers,amsmath,amssymb]{revtex4}

\usepackage{graphicx}
\usepackage{bm}
\usepackage{color}
\def\slashchar#1{\setbox0=\hbox{$#1$}
   \dimen0=\wd0 \setbox1=\hbox{/} \dimen1=\wd1
   \ifdim\dimen0>\dimen1 \rlap{\hbox to \dimen0{\hfil/\hfil}} #1
   \else  \rlap{\hbox to \dimen1{\hfil$#1$\hfil}} / \fi}
\begin{document}
\title{Medium effects in the deep-inelastic charged lepton/neutrino-A scattering}
\author{H. Haider$^1$, F. Zaidi$^1$, M. Sajjad Athar$^1$, S. K. Singh$^1$ and I. Ruiz Simo$^2$}
\address{$^1$ Department of Physics, Aligarh Muslim University, Aligarh-202 002, India\\
$^2$ Departamento de F\'{\i}sica At\'omica, Molecular y Nuclear,
and Instituto de F\'{\i}sica Te\'orica y Computacional Carlos I,
Universidad de Granada, Granada 18071, Spain} 

\begin{abstract}
In this work, we have discussed the recent developments that have taken place to understand the differences 
 in the weak $F_{2A}^{Weak} (x,Q^2)$ and 
 electromagnetic $F_{2A}^{EM} (x,Q^2)$  nuclear structure functions. Also we present the results of our work on nuclear medium effects on $F_{2A}^{Weak} (x,Q^2)$ and 
 $F_{2A}^{EM} (x,Q^2)$ for a wide range of $x$ and $Q^2$. These results have been obtained using a microscopic nuclear model, where to incorporate 
 nuclear medium effects, Fermi motion, binding energy, nucleon correlations, mesonic contributions from pion and rho mesons and shadowing effects are considered.
 The calculations are performed in local density approximation using relativistic nucleon spectral function. We have also compared the theoretical results with the recent experimental data
on electromagnetic and weak structure functions. Furthermore, we have studied the nuclear medium effects in Drell-Yan(DY) process and present the results for differential cross section, and the results are compared with the data of E772 experiment.
\end{abstract}

\maketitle

\section{Introduction}
   The inclusive cross sections for the charged leptons or (anti)neutrino induced reactions on nucleons and nuclei, at 
  intermediate energies are expressed in terms of structure functions corresponding to excitations of various resonances 
  like $\Delta,~N^\ast$, etc., lying in the first or higher resonance region depending on the center of mass (CM) energy W of the final hadrons, while 
  at high energies and $Q^2$,  the inclusive cross sections are expressed in terms of the structure functions corresponding to the DIS processes.
  The DIS region is understood to be the kinematic region with the center of mass energy W $\ge$ 2.0 $GeV$ and 
    $Q^2 \ge$ 1.0 $GeV^2$. 
    Recently JLab has performed experiments using continuous electron beam facility with 
  energies in the range of approximately 2-6 $GeV$, and precise measurements have 
  been performed for the structure functions $F_{1N}(x,Q^2)$, $F_{2N}(x,Q^2)$, longitudinal structure function $F_{LN}(x,Q^2)$
  extending to low $Q^2$($<5~GeV^2$) using several nuclear targets~\cite{Mamyan:2012th}.
  The modification of structure function in nuclear medium has also been studied earlier by NMC~\cite{Arneodo:1997}, JLab~\cite{Aubert:1986yn}, 
  SLAC~\cite{Gomez:1993ri}, BCDMS~\cite{Benvenuti:1994bb}, etc. collaborations, in some nuclear targets. JLab also plans to upgrade electron beam energy to 12 GeV and perform 
  the experiments using several nuclear targets. JLab experiment, has observed that Bloom-Gilman duality 
  to work even at $Q^2 \sim 1~GeV^2$ or may be less than $1~GeV^2$, which means that resonance structure function averaged 
  over the scaling variable is almost equal to the 
  deep inelastic structure function. In the weak sector, several experiments have been performed~\cite{Berge:1989hr,Oltman:1992pq,Tzanov:2005kr} and 
  few others are going on to study neutrino
oscillation physics while some of them like MINER$\nu$A~\cite{Mousseau:2016snl} and DUNE~\cite{Strait:2016mof} are specially designed to precisely measure
neutrino and antineutrino cross sections in the DIS region on some nuclear targets. 
One of the major source of systematic errors in all the neutrino oscillation experiments being performed in the few GeV energy region
($1 < E_\nu < 5$ $GeV$) 
 arises due to lack of the understanding of nuclear medium effects(NME) in the neutrino/antineutrino-nucleus scattering cross section. 
 There are many theoretical calculations of NME in Quasielastic(QE), Inelastic(IE) scattering and deep-inelastic scattering(DIS)
 processes which give the 
 uncertainty bands in the neutrino-nucleus cross sections arising due to NME.
 Many efforts have been made in the last one decade to understand medium effects in the quasielastic and one pion production processes arising from 
 neutrino/antineutrino interactions from nuclear targets, not many attempts have been made to understand medium effects in the DIS region.
 Despite of these efforts, the cross sections have still an uncertainty
 of 20-30$\%$ and a better understanding of NME would play decisive role in the study of CP violation in the leptonic sector or in very precise determination 
 of oscillation parameters. To understand NME in the DIS region, two different approaches  are adopted, one is 
 a phenomenological approach and the other is a theoretical approach. In the phenomenological analysis there are few approaches for determining nuclear PDFs. 
 In most PDF analyses, the nuclear correction factors were taken from lepton-nucleus and p-nucleus scattering data, some of them also include Drell-Yan data as well as 
(anti)neutrino-nucleus scattering data. Most of the studies like of Eskola et al.~\cite{Eskola:1998df,Eskola:1998iy,Eskola:2009uj} and de Florian et al.~\cite{deFlorian:2011fp} do not find 
 a difference in the nuclear correction factor obtained using $l^\pm-A$ and $\nu_l/\bar\nu_l-A$ scattering data. However, recent studies by 
 CTEQ-Grenoble-Karlsruhe 
collaboration (nCTEQ)~\cite{Kovarik:2010uv} have shown that the nuclear correction factor is different in electromagnetic(EM) interaction vs weak interaction.
Therefore, it is important to understand the nature of differences in electromagnetic and weak structure functions as well as to understand the violation of Callan-Gross relation in nuclei due to NME.
  
\section{Formalism}
The basic reaction for charged/neutral lepton induced DIS process on bound nucleons, in nuclear target is given by
\begin{eqnarray}
 l^\pm(k) +N(p) &\to& l^\pm(k') + X(p')\nonumber\\
 \nu_l/\bar\nu_l(k) +N(p) &\to& l^-/l^+(k') + X(p')
\end{eqnarray}
for which the leptonic tensor 
\begin{equation}
L_{\alpha\beta}
= 2 \left( \underbrace{k_\alpha k'_\beta + k'_\alpha k_\beta - g_{\alpha\beta} k \cdot k'}_{symmetric} \pm \underbrace{  i \epsilon_{\alpha\beta\rho\sigma}k^\rho k'^\sigma}_{antisymmetric} \right)\ ,
\label{eq:Lmunu}
\end{equation}
and the nuclear hadronic tensor $W^{\alpha\beta}_A$ is 
\begin{eqnarray}
W^{\alpha\beta}_A
&=& W_{1A}(\nu,Q^2)
  \left( { q^\alpha q^\beta \over q^2} - g^{\alpha\beta} \right)
+ { W_{2A}(\nu ,Q^2) \over M_A^2 }
  \left( p_A^\alpha - { p_A\cdot q \over q^2 } q^\alpha \right)
  \left( p_A^\beta - { p_A\cdot q \over q^2 } q^\beta \right)\nonumber\\
  &&-\underbrace{{i \over 2M_A^2} \epsilon^{\alpha\beta\rho\sigma} p_{A \rho} q_\sigma~ W_{3A}(\nu,Q^2)}_{parity~ violating}\ ,
\label{eq:WAmunu_weak}
\end{eqnarray}
which is expressed in terms of nuclear structure functions $W_{iA}(\nu,Q^2)$ where $i=1,2$ for EM interaction and $i=1,2,3$ for weak interaction.
In the case of electromagnetic interaction the antisymmetric term of leptonic tensor(Eq.\ref{eq:Lmunu}) and the parity violating term of hadronic tensor(Eq.\ref{eq:WAmunu_weak})
do not contribute.

These structure functions are expressed in terms of dimensionless nuclear structure functions as
\begin{eqnarray}
 F_{1A}(x) &=& M_A W_{1A}(\nu,Q^2)\ ;	F_{2A}(x) = \nu W_{2A}(\nu,Q^2)\ ;  F_{3A}(x) = \nu W_{1A}(\nu,Q^2).
\end{eqnarray}
We have performed the numerical calculations in the local density approximation and obtained the nuclear hadronic tensor in terms of hole spectral function($S_h$) and 
nucleon hadronic tensor using which the expression of nuclear structure functions for nonisoscalar nuclear target 
are obtained as~\cite{SajjadAthar:2009cr,Haider:2011qs,Haider:2012nf,Haider:2014iia,Haider:2015vea,Haider:2016zrk}
\begin{eqnarray}	\label{conv_WA1}
F_{1A}^{EM,Weak}(x_A, Q^2) &=& 2\sum_{\tau=p,n} AM_N \int \, d^3 r \, \int \frac{d^3 p}{(2 \pi)^3} \, 
\frac{M_N}{E ({\bf p})} \, \int^{\mu_\tau}_{- \infty} d p_0 S_h^\tau (p_0, {\bf p}, \rho^\tau(r)) \times \nonumber\\
&&\left[\frac{F_{1}^{\tau, EM,Weak}(x_N, Q^2)}{M_N}+ \frac{{p_x}^2}{M_N^2} \frac{F_{2}^{\tau, EM,Weak}(x_N, Q^2)}{\nu_N}\right],
\end{eqnarray}
 \begin{eqnarray} \label{em_f2_noniso}
F_{2A}^{EM,Weak}(x_A,Q^2)  &=&  2\sum_{\tau=p,n} \int \, d^3 r \, \int \frac{d^3 p}{(2 \pi)^3} \, 
\frac{M_N}{E ({\bf p})} \, \int^{\mu_\tau}_{- \infty} d p_0 S_h^\tau (p_0, {\bf p}, \rho^\tau(r)) \times \nonumber \\
&&\left[\frac{Q^2}{q_z^2}\left( \frac{|{\bf p}|^2~-~p_{z}^2}{2M_N^2}\right) +  \frac{(p_0~-~p_z~\gamma)^2}{M_N^2} \left(\frac{p_z~Q^2}{(p_0~-~p_z~\gamma) q_0 q_z}~+~1\right)^2\right]~\times\nonumber\\
&& \left(\frac{M_N}{p_0~-~p_z~\gamma}\right) ~F_2^{\tau, EM,Weak}(x_N,Q^2),   
\end{eqnarray}
where $M_N$ is the nucleon mass, $x_A=\frac{Q^2}{2 A M_N \nu}$, $\rho(r)$ is the nuclear density, $\gamma=\sqrt{1+{4 M_N^2 x^2 \over Q^2}}$,
$S_h$ is the hole spectral function
which takes into account the effect of Fermi motion, binding energy and nucleon correlations, the expression for which is taken from Ref.~\cite{FernandezdeCordoba:1991wf}and $F_{2}^{\tau, EM,Weak}(x_N,Q^2)$ is the nucleon structure function which are expressed in terms of parton distribution
functions(PDFs).
The only difference in obtaining the expressions of weak and EM nuclear structure functions 
comes from the nucleon PDFs and intermediate boson self energy. In a similar way, the expression of weak nuclear
structure function $F_{3A}^{Weak}(x_A, Q^2)$ may be obtained~\cite{Haider:2011qs}.
The detailed discussion of formalism is given in Refs.~\cite{SajjadAthar:2009cr,Haider:2011qs,Haider:2012nf,Haider:2014iia,Haider:2015vea,Haider:2016zrk}. 

In the numerical calculations we have included the contribution from mesonic cloud for which the structure functions may be obtained by following the same procedure as 
in the case of nucleons except that nucleon propagator is now replaced by the meson propagator:
\begin{eqnarray}\label{meson_f1}
 F_{1 A,i}^{ EM,Weak}(x,Q^2)  &=& -6\times \kappa \times A M_N \int \, d^3 r \, \int \frac{d^4 p}{(2 \pi)^4} \, 
        \theta (p_0) ~\delta I m D_i (p) \;2m_i~\nonumber\\
     &\times&  \left[\frac{F_{1i}^{EM,Weak}(x_i)}{m_i} +\frac{|{\bf p}|^2-p_z^2}{2(p_0q_0-p_zq_z)}\frac{F_{2i}^{ EM,Weak}(x_i)}{m_i} \right],
\end{eqnarray}
\begin{eqnarray} \label{pion_f21}
F_{2 A,i}^{ EM,Weak }(x,Q^2)  &=&  -6\times \kappa \int \, d^3 r \, \int \frac{d^4 p}{(2 \pi)^4} \, 
        \theta (p_0) ~\delta I m D_i (p) \;2m_i~\left(\frac{m_i}{p_0~-~p_z~\gamma}\right)\times ~F_{2i}^{ EM,Weak}(x_i)\times\nonumber \\
&&\left[\frac{Q^2}{(q_z)^2}\left( \frac{|{\bf p}|^2~-~(p_{z})^2}{2m_i^2}\right)  
+  \frac{(p_0~-~p_z~\gamma)^2}{m_i^2} \left(\frac{p_z~Q^2}{(p_0~-~p_z~\gamma) q_0 q_z}~+~1\right)^2\right] ~,
\end{eqnarray}
where $i=\pi~or~\rho$, $x_i=\frac{Q^2}{-2 p \cdot q}$, $m_i$ is the mass of meson and the constant factor $\kappa$ is 1 in the case of pion and 2 in the case of $\rho$ meson~\cite{Haider:2015vea}.
For the pionic contribution we have used the PDFs given by Gluck et al.~\cite{Gluck:1991ey} and the same PDFs are used for the $\rho$ meson contribution. Furthermore, we have included the effect of shadowing following the works of Kulagin and Petti~\cite{Kulagin:2004ie}. 

\section{Results and Discussion}
\begin{figure}
\begin{center}
  \includegraphics[height=4 cm, width=6.5 cm]{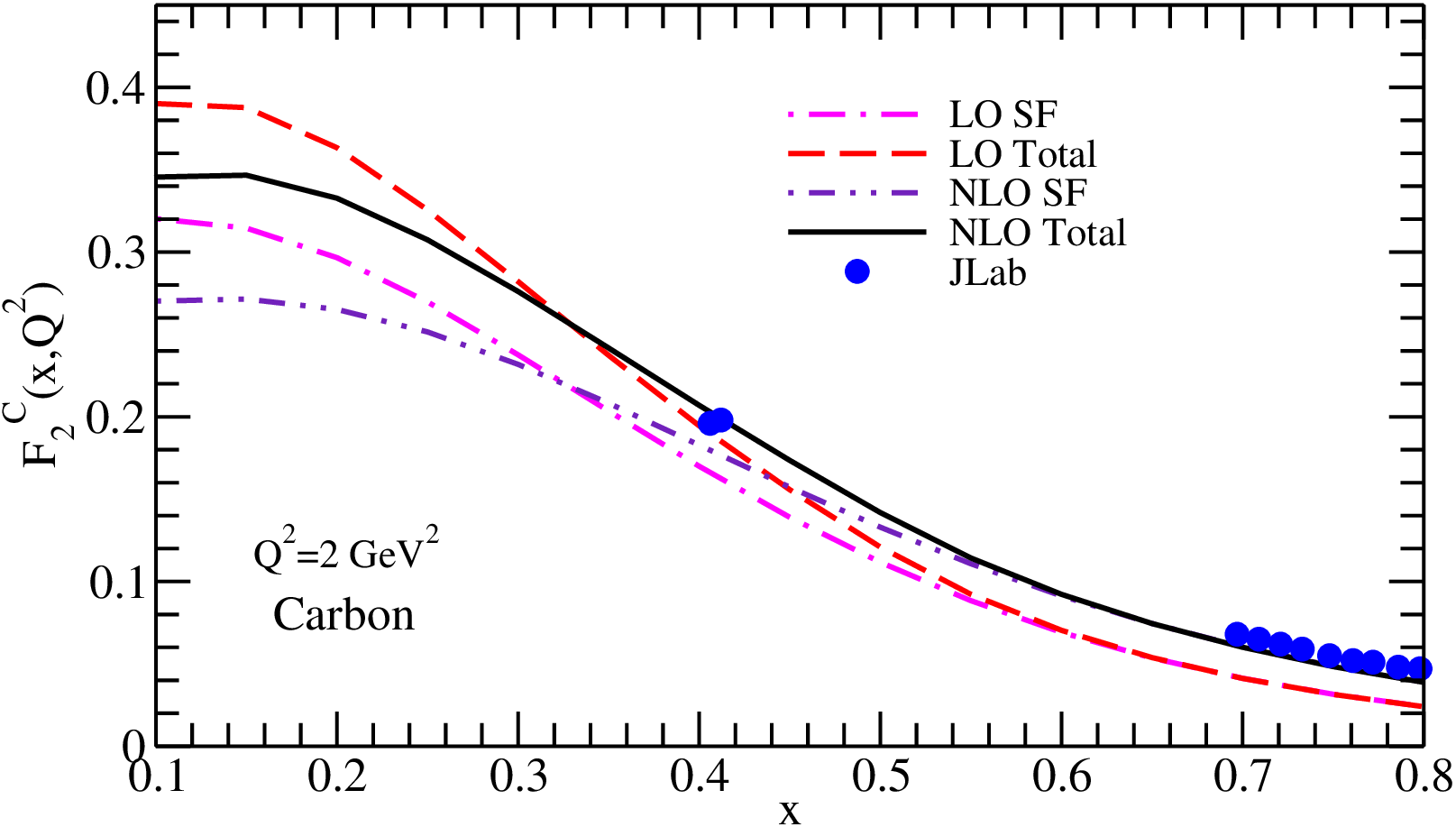}
 \includegraphics[height=4 cm, width=6.5 cm]{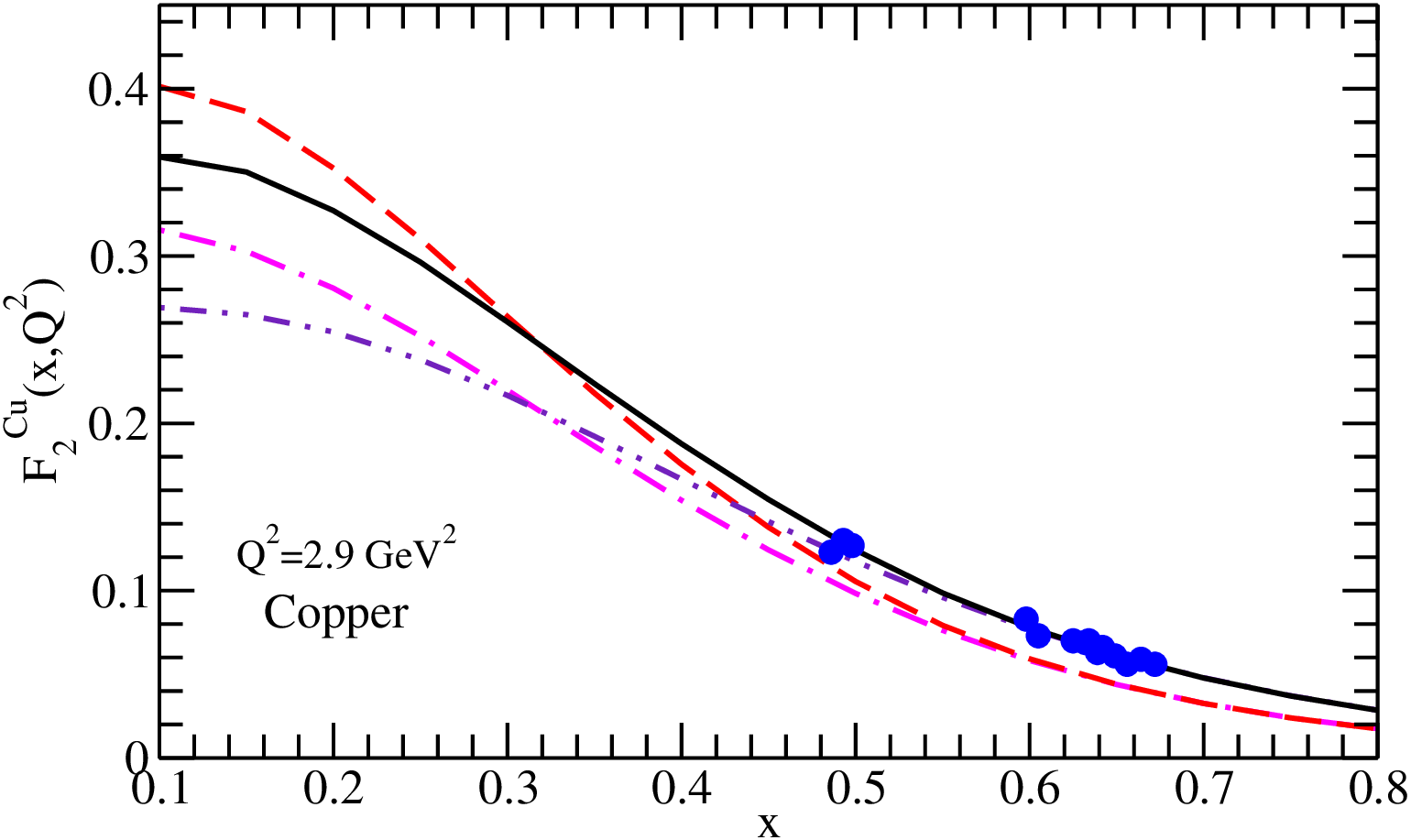}
\end{center}
\caption{Results of nuclear structure function for EM interaction in carbon and copper at both LO and NLO for a fix $Q^2$ are shown and compared with 
 JLab data[1].}
 \label{fig1}
\end{figure}
\begin{figure}
\begin{center}
  \includegraphics[height=4 cm, width=15 cm]{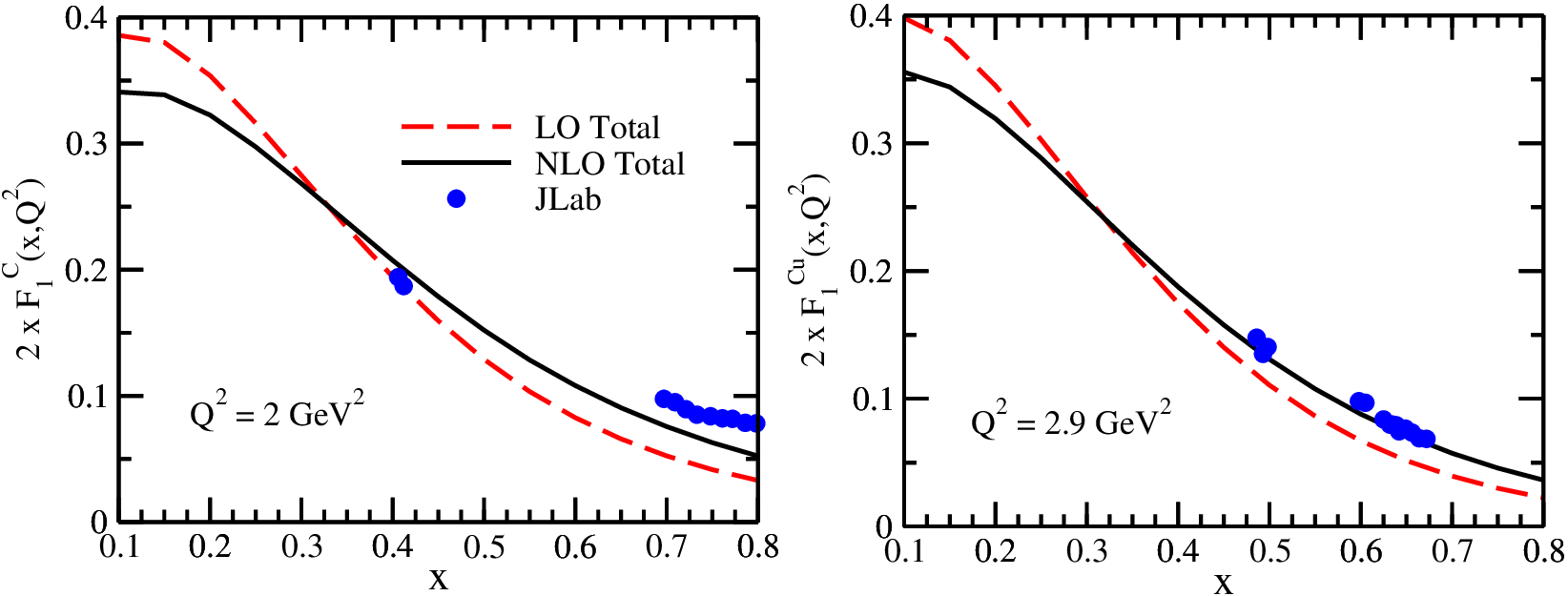}
\end{center}
\caption{Results of nuclear structure function for EM interaction in carbon and copper at a fixed value of
 $Q^2$ are shown and are compared with the JLab data[1].}
 \label{fig2}
\end{figure}

In Figs.\ref{fig1} and \ref{fig2}, we have presented the results for electromagnetic nuclear structure functions $F_{2}^A(x,Q^2)$ and $2 x F_{1}^A(x,Q^2)$, 
 in $^{12}C$ and $^{63}Cu$ nuclear targets at a fixed value of $Q^2$. In Fig.\ref{fig1}, the curves depict 
 the results that are obtained using spectral function(SF) and full model(SF+mesonic contribution+shadowing effect) at the leading order(LO) 
 and next-to-leading order(NLO). 
 We find that the results at LO with spectral function is about 18$\%$ smaller at x=0.2 in comparison to the
results obtained using the full model at $Q^2=2~GeV^2$. 
This difference decreases with the increase in x, for example at x=0.4, it is 12$\%$ and becomes
 almost negligible at x=0.6. When the results obtained by using the full 
  model at NLO are compared with the results evaluated at LO, we find that there is decrease in the results 
  from the LO values. Our results at NLO with full model are in very good agreement with the experimental data of JLab.

In Fig.\ref{fig2}, the results of $2 x F_{1}^A(x,Q^2)$ are qualitatively similar in nature as found in the case of $F_2^A(x,Q^2)$, 
however, some quantitative difference 
in the region of low x where mesonic effects are dominant is found. These results are also 
compared with the data of JLab experiment~\cite{Mamyan:2012th} and are in reasonable agreement with it. 
\begin{figure}
\begin{center}
  \includegraphics[height=4 cm, width=6.5 cm]{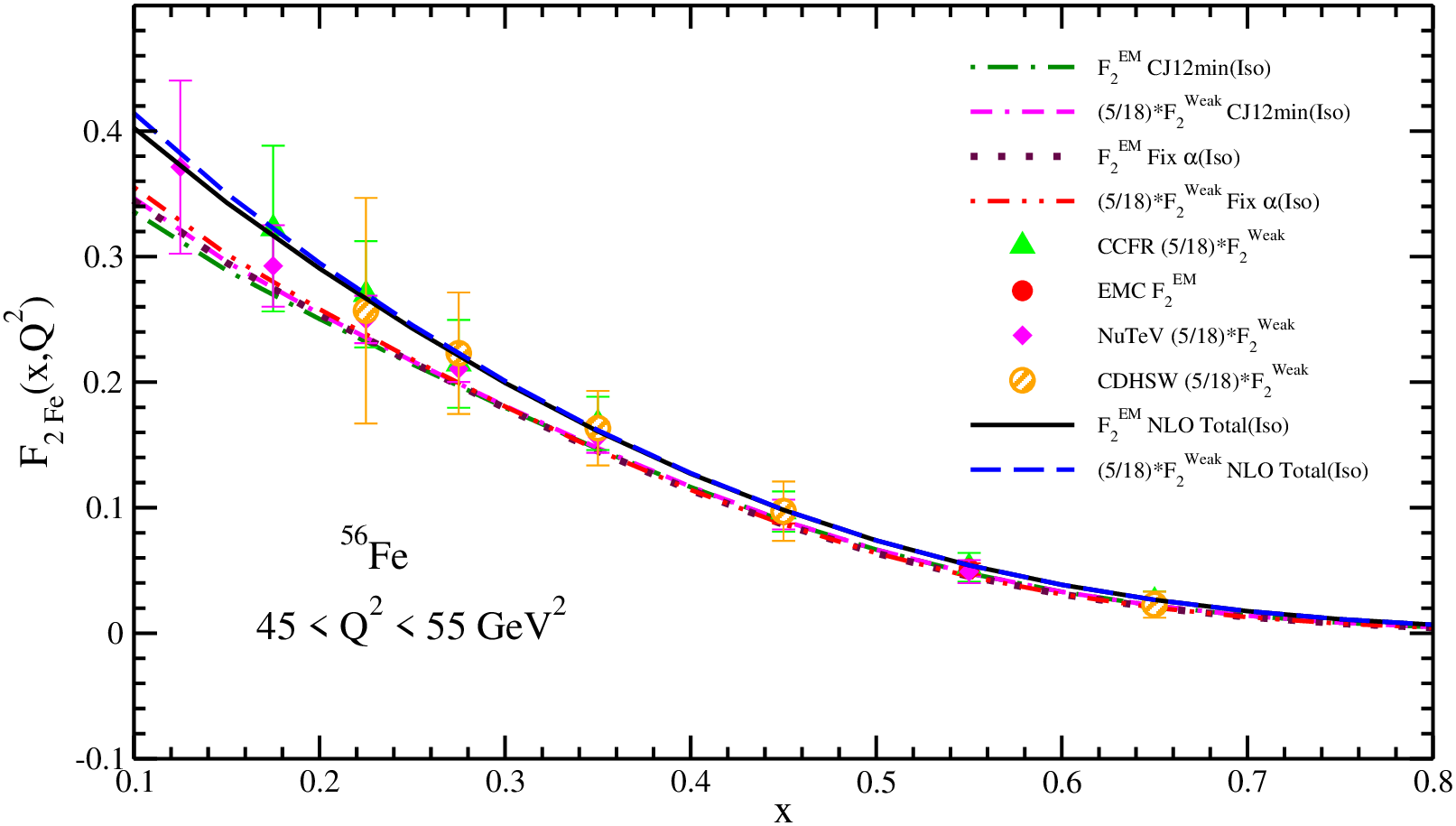}
    \includegraphics[height=4 cm, width=6.5 cm]{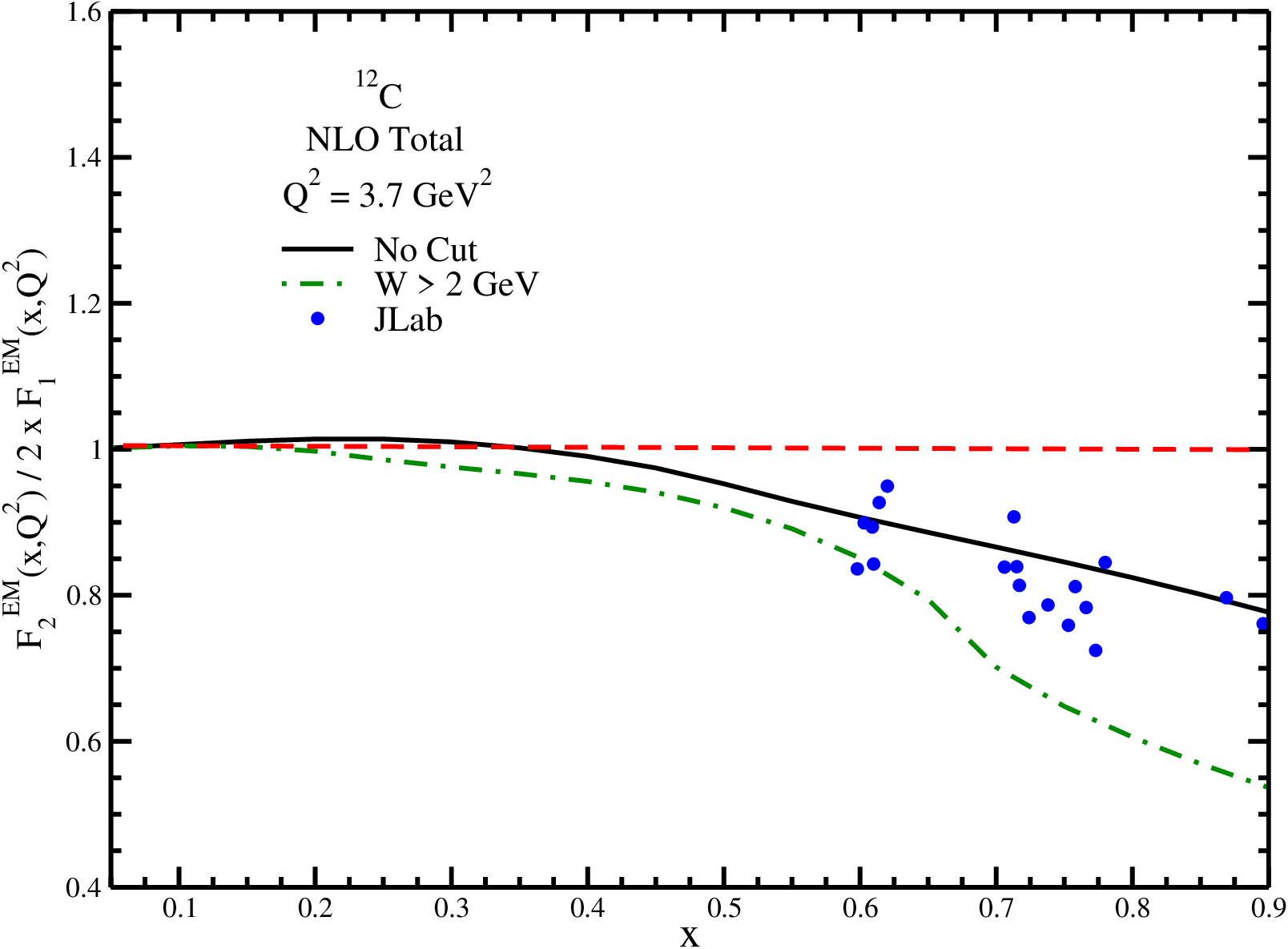}
\end{center}
\caption{{\bf (i)Left panel:} Nuclear structure function for electromagnetic and weak interaction in iron and are compared with the experimental data.
{\bf (ii) Right panel:} Results for the ratio of $F_{2A}^{EM}(x,Q^2) \over 2 x F_{1A}^{EM}(x,Q^2) $ showing the deviation of Callan-Gross relation 
in carbon at a fixed value of $Q^2$ and are compared with the JLab data[1].}
 \label{fig3}
\end{figure}

In Fig.\ref{fig3}(Left panel), we have shown the results for weak nuclear structure functions and compared them with the results obtained in the case of EM interaction.
The results are also presented by using the nuclear PDFs CJ12min~\cite{Owens:2012bv} for fix value of strong coupling constant as well as for $Q^2$ evolution. 
From the figure one may observe that the results obtained with nucleon PDFs
CTEQ6.6 are different than the results obtained using nuclear PDFs CJ12min. 
It is also noticeable that at low $x$, EM structure 
 function is slightly lower than the weak structure function which is about $\sim 4\%$ in iron 
 at $x=0.1$, and for higher values of $x~>~0.3$ it becomes almost negligible. 
 However, for the heavier nuclear targets the difference in EM and weak structure functions is found to be large. 
 
 In the right panel of Fig.\ref{fig3}, the ratio of nuclear structure functions $F_{2A}^{EM}(x,Q^2) \over 2 x F_{1A}^{EM}(x,Q^2) $ is shown for 
 carbon at $Q^2=3.7~GeV^2$. It may be observed that
 the ratio is less than unity and further suppressed when we apply a cut of 2 $GeV$ on CM energy $W$. Hence, from the results it may be
 concluded that Callan-Gross(CG) relation deviates inside the nuclear medium.
 These results are also in agreement with the JLab data~\cite{Mamyan:2012th}.
\begin{figure}
\begin{center}
  \includegraphics[height=4 cm, width=6.5 cm]{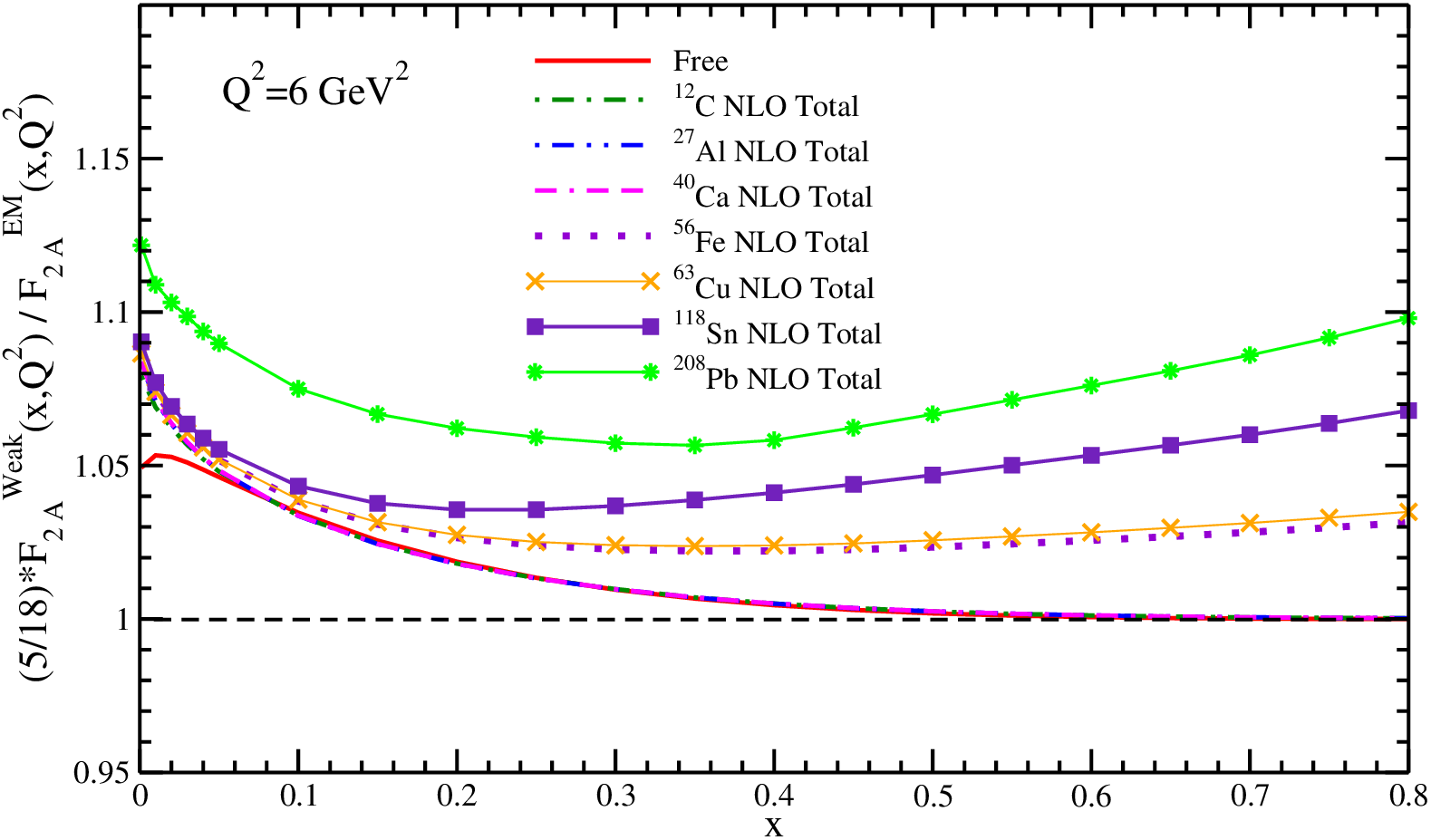}
    \includegraphics[height=4 cm, width=6.5 cm]{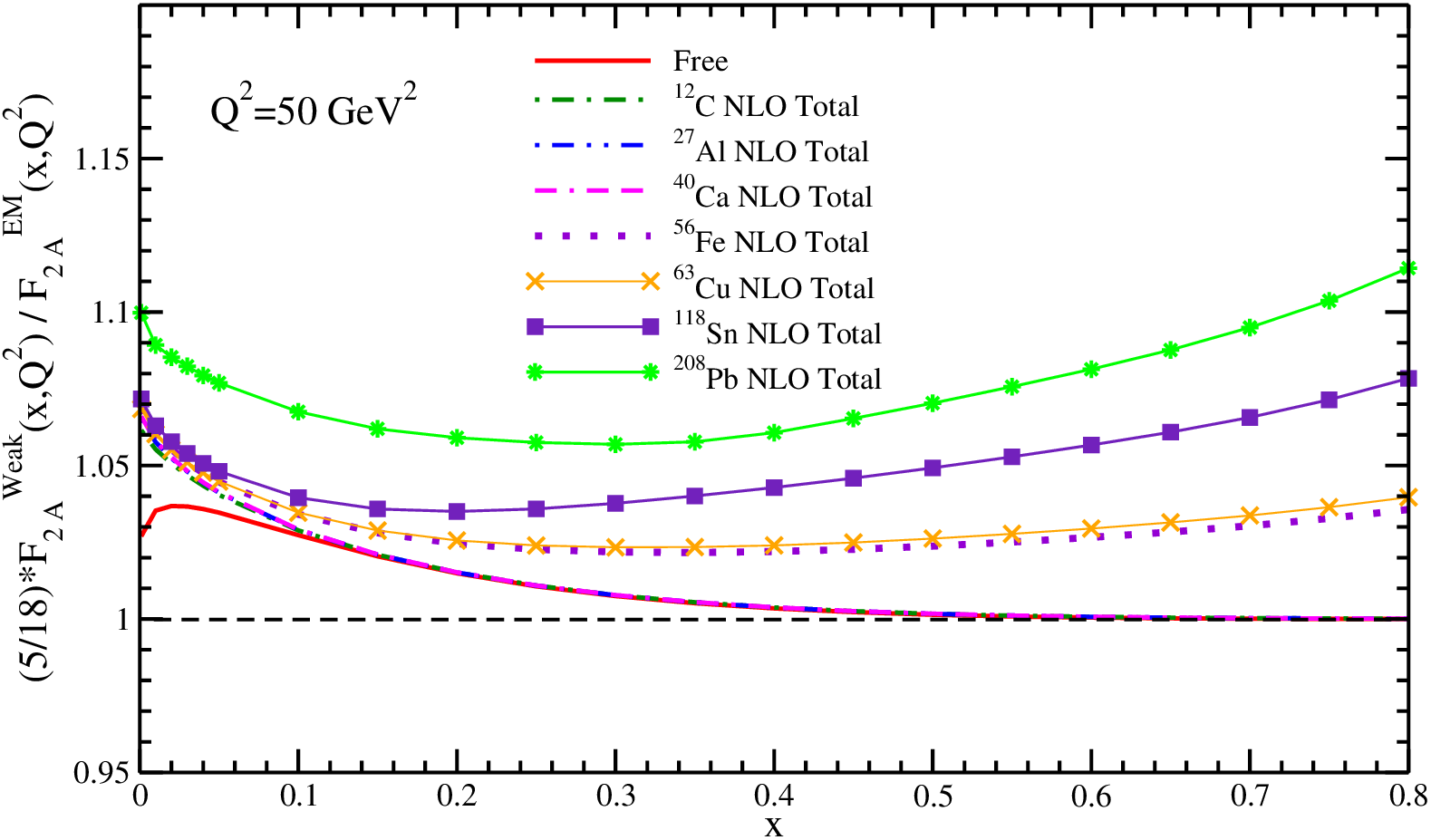}
\end{center}
\caption{Results for the ratio of nuclear structure functions $\frac{\frac{5}{18} F_{2A}^{Weak}(x,Q^2)}{F_{2A}^{EM}(x,Q^2)}$ 
are shown in different nuclear targets at $Q^2=6, ~50~GeV^2$
using the full model at NLO.}
 \label{fig4}
\end{figure}

 In Fig.~\ref{fig4}, we show the results for the ratio $\frac{\frac{5}{18} F_{2A}^{Weak}(x,Q^2)}{F_{2A}^{EM}(x,Q^2)}$ in 
 various nuclear targets like 
 $^{12}C$, $^{27}Al$, $^{40}Ca$, $^{56}Fe$, $^{63}Cu$, $^{118}Sn$ and $^{208}Pb$ at $Q^2=6,~50~GeV^2$. It may be noticed from the figure that
 nonisoscalarity effect is larger for heavier nuclear targets like $^{118}Sn$ and $^{208}Pb$ which implies that the difference in charm and strange 
 quark distributions could be
 significant for heavy nuclei. 
Furthermore, the ratio is found to be $x$ as well as $Q^2$ dependent.
\begin{figure}
\begin{center}
    \includegraphics[height=5 cm, width=12 cm]{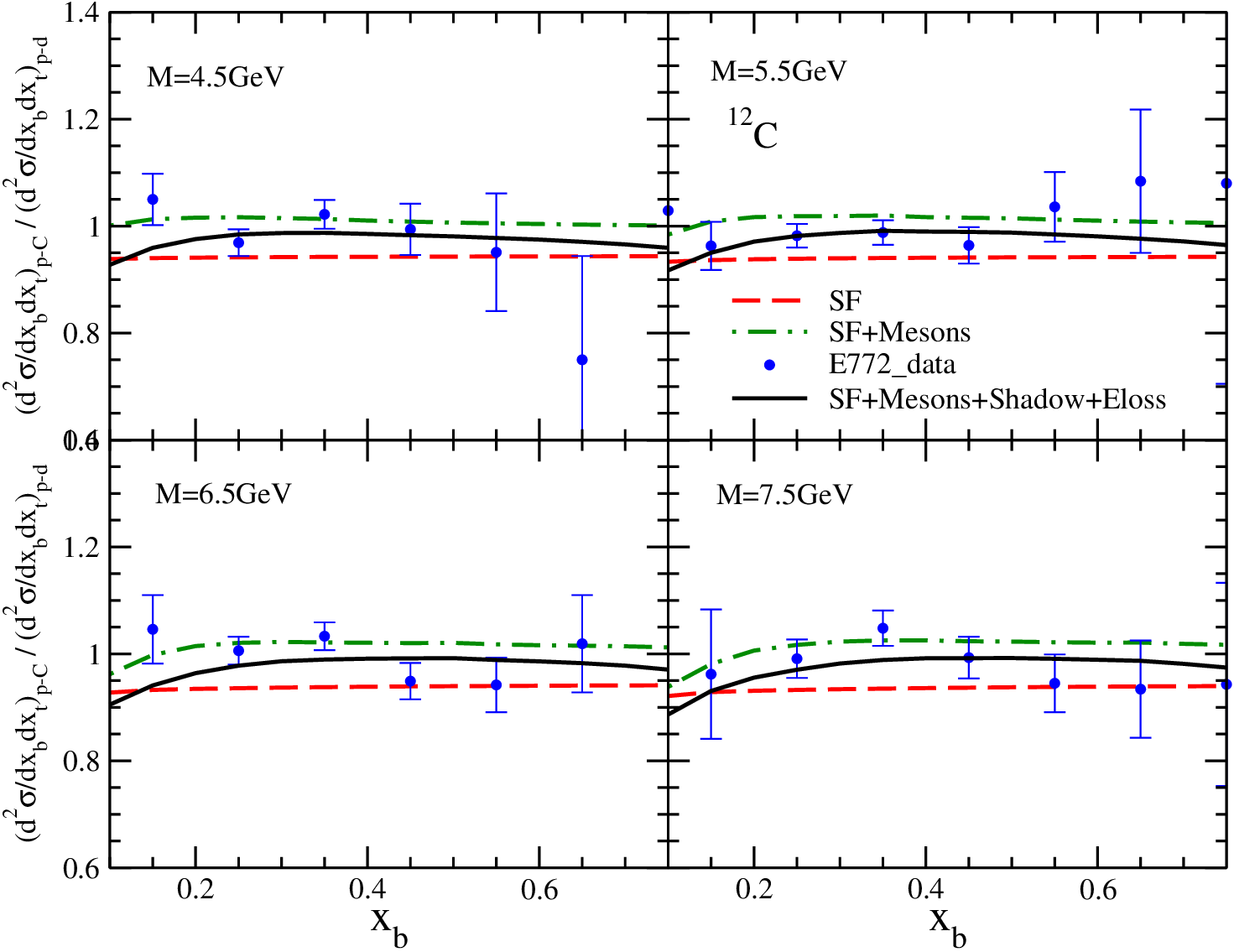}
\end{center}
\caption{$\frac{\left(\frac{d\sigma}{dx_b~dx_t}\right)_{p - ^{^{12}}C}}{\left(\frac{d\sigma}{dx_b~dx_t}\right)_{p - ^{^2}D}}$ vs $x_b$ at
 E=800GeV($\sqrt{s_N}$=38.8GeV). The results in the different columns are obtained at different values of 
 M(=$\sqrt {q^2}$). Experimental points are data of E772 experiment[27,28].}
 \label{fig5}
\end{figure}

In Fig.\ref{fig5}, we present the results for the ratio 
 ${\it R}=\frac{\left(\frac{d\sigma}{dx_b~dx_t}\right)_{p - A}}{\left(\frac{d\sigma}{dx_b~dx_t}\right)_{p - ^{^2}D}}$
  vs $x_b$, where numerator of the ratio stands for the proton-nucleus and denominator stands for
  the proton-deuteron DY differential cross section for which the expression are given in Ref.~\cite{Haider:2016tev}. 
  The results are obtained for M=4.5, 5.5, 6.5 and 7.5 $GeV$ in carbon at center of mass energy $\sqrt{s_N}= 38.8~GeV$ and compared 
  with the E772 data~\cite{alde,dyhepdata}. We find that the nuclear structure effects 
  due to bound nucleon lead to a suppression in the DY yield of about  $5-6\%$ in $^{12}C$ in the region 
   of $0.2~<~x_b~<0.6$.  Furthermore, we find that there is a significant contribution of mesons which increases the DY ratio i.e. its effect 
   is opposite to the effect 
  of spectral function. For example, the DY yield increases by around $6-8\%$ for $0.2~<~x_b~<0.6$ in $^{12}C$. Moreover, we observe that the effect is more at 
  low $x_b$($\sim 0.2-0.3$) than at high $x_b$.  
  We find the contribution from rho meson cloud to be much smaller than the contribution from pion cloud.   
  When the shadowing corrections are included there is further suppression in the DY yield and it is effective in the low region of $x_b(\le~0.2)$.
  The effect of beam energy loss is also to reduce the DY yield. 
  Both effects add to the suppression obtained using spectral function, 
  where as the mesonic effects lead to an enhancement. The net effect of shadowing and the energy loss effect is to reduce the DY yield by about  
  $7\%$  at $x_b=0.1$ in $^{12}C$ which becomes $4\%$  at $x_b=0.2$ for $M=4.5~GeV$. 
\section{Conclusions}
From the present study we may conclude the following:
\begin{enumerate}
\item The nuclear structure functions $F_{2A}(x,Q^2) $ and $2 x F_{1A}(x,Q^2)$ are different due to nuclear medium effects and as a consequence of which
Callan-Gross relation deviates inside the nuclear medium.
\item Due to the nuclear medium effects, we found the difference in $F_{2A}^{EM}(x,Q^2)$ and $F_{2A}^{Weak}(x,Q^2)$ structure functions 
in the region of low $x$ which vanishes for high x values.
\item We have also found the difference between the nuclear structure functions obtained by using nucleon and nuclear parton distribution functions.
 \item In the case of Drell-Yan process both the reduction as well as the enhancement due to the nuclear structure effect, shadowing and mesonic cloud contribution,
 are found to be of similar magnitude as in the case of DIS of charged leptons.
 \end{enumerate}

\begin{footnotesize}

\end{footnotesize}

\end{document}